\def\e{{\rm e}}
\def\be{\begin{equation}}
\def\ee{\end{equation}}
\def\bea {\begin{eqnarray}}
\def\eea {\end{eqnarray}}
\documentstyle[prl,aps,twocolumn]{revtex}

\begin{document}
\input{epsf.tex}
\wideabs{
\title{Thermodynamics of the Spin-1/2 Antiferromagnetic
Uniform Heisenberg Chain}
\author{ A. Kl\"umper$^{1,2}$ and D. C. Johnston$^3$}
\address{$^1$Institut f\"ur Theoretische Physik, Universit\"at zu K\"oln,
Z\"ulpicher Strasse 77, 50937 K\"oln, Germany}
\address{$^2$Fachbereich Physik, Universit\"at Dortmund, 44221
Dortmund, Germany}
\address{$^3$Ames Laboratory and Department of Physics and Astronomy, Iowa
State University, Ames, Iowa 50011}
\date{Revised Manuscript Submitted to Physical Review Letters on 28
January 2000}
\maketitle
\begin{abstract}\hglue 0.15in
  We present a new application of the traditional thermodynamic Bethe
  ansatz to the spin-1/2 antiferromagnetic uniform Heisenberg chain
  and derive exact nonlinear integral equations for just {\em two}
  functions describing the elementary excitations. Using this approach
  the magnetic susceptibility $\chi$ and specific heat $C$ versus
  temperature $T$ are calculated to high accuracy for
  $5\times10^{-25}\leq T/J\leq 5$. The $\chi(T)$ data agree very well
  at low $T$ with the asymptotically exact theoretical low-$T$
  prediction of S. Lukyanov, Nucl.\ Phys.\ B {\bf 522}, 533 (1998).
  The unknown coefficients of the second and third lowest-order
  logarithmic correction terms in Lukyanov's theory for $C(T)$ are
  estimated from the $C(T)$ data.
\end{abstract}
}

The spin $S = 1/2$ antiferromagnetic (AF) uniform Heisenberg chain has a
long and distinguished history in condensed matter physics and exhibits
unusual static and dynamical properties unique to one-dimensional
spin systems.  It has been used as a testing ground for many theoretical
approaches.  The Hamiltonian is ${\cal H} = J \sum_{i}
\bbox{S}_i\cdot\bbox{S}_{i+1}$, where $J>0$ is the AF Heisenberg exchange
interaction between nearest-neighbor spins.  In this paper we usually set
$k_{\rm B} = 1$ and $g \mu_{\rm B} = 1$ where $k_{\rm B}$ is Boltzmann's
constant, $g$ is the spectroscopic splitting factor of the spins and
$\mu_{\rm B}$ is the Bohr magneton; also, the reduced temperature
$t\equiv T/J$ where $T$ is the absolute temperature.

The $S = 1/2$ Heisenberg chain is known to be exactly
solvable\cite{Bethe31}, i.e.\ all eigenvalues can be obtained from the
so-called Bethe ansatz equations.  Despite the amazing property of being
integrable, the Heisenberg chain has defied many attempts to calculate
physical observables including thermodynamic quantities. A rather direct
evaluation of the partition function was constructed in\cite{TakTBA} and
is known as the ``thermodynamic Bethe ansatz'' (TBA),
but this did not allow for
high accuracy calculations especially in the low temperature region. The
fundamental problem in \cite{TakTBA} is the necessity to deal with
infinitely many coupled nonlinear integral equations for which the
truncation procedures are difficult to control.

The possibility to accurately calculate the physical properties of the
$S = 1/2$ Heisenberg chain improved following the development of the
path integral formulation of the transfer matrix treatment of quantum
systems\cite{SuzukiI87}.  On the basis of a Bethe ansatz
solution\cite{Tak91} to the quantum transfer matrix, Eggert, Affleck
and Takahashi in 1994 obtained numerically exact results for the
magnetic susceptibility $\chi(t)$ down to much lower temperatures than
before and compared these with their low-$t$ results from conformal
field theory\cite{Eggert1994}.  They found, remarkably, that
$\chi(t\to 0)$ has infinite slope: their conformal field theory
calculations showed that the leading order $t$ dependence is
$\chi(t\to 0) = \chi(0)\{1 + 1/[2\ln(t_0/t)]\}$, where the value of
$t_0$ is not predicted by the field theory.  Such log terms are called
``logarithmic corrections''.  From their comparison of their field
theory and Bethe ansatz calculations which extended down to $t =
0.003$, Eggert, Affleck and Takahashi estimated $t_0 \approx
7.7$\cite{Eggert1994}. Their numerical $\chi(t)$ values are up to
$\sim 10$\% larger than the former Bonner-Fisher \cite{Bonner1964}
extrapolation for $t\lesssim 0.25$.

Lukyanov has recently presented an exact asymptotic field theory for
$\chi(t)$ and the specific heat $C(t)$ at low $t$, including the exact
value of $t_0$ \cite{Lukyanov1997}. These results are claimed to be exact
in the sense of a renormalization group treatment close to a fixed point
where only few operators are responsible for perturbations.  Questions
arising about such calculations are whether these operators have been
correctly identified and whether the effective theory has been properly
evaluated.  A meaningful test of Lukyanov's theory is only possible using
numerical data of very high accuracy and at extremely low temperatures,
such as we have attained in our numerical calculations to be presented
below.

In this Letter we present a new application of the traditional TBA to the
spin-1/2 Heisenberg chain and derive exact nonlinear integral equations
[Eqs.~(\ref{NLIE}--\ref{kernel}) below] involving just {\em two}
functions describing the elementary excitations.  Our derivation evolved
from earlier work by one of us using the powerful lattice
approach\onlinecite{Klumper1993,Klumper1998}. By means of a lattice path
integral representation of the finite temperature Heisenberg chain and
the formulation of a suitable quantum transfer matrix, a set of
numerically well-posed expressions for the free energy was derived.  A
serious disadvantage of this approach lies in the complicated and
physically non-intuitive mathematical constructions, which strongly
inhibits generalizations to other integrable, notably itinerant fermion
models.  The present work is a new analytic derivation of the
finitely-many integral equations of \cite{Klumper1993,Klumper1998} by
means of the intuitive TBA approach.  Our Eqs.~(\ref{NLIE}--\ref{kernel})
are identical to those obtained in \cite{Klumper1993} by a rigorous,
however much more involved method.  In our new construction, we assume
that magnons (on paths $C_\pm$) are elementary excitations and contain
all information about the thermodynamics.  Bound states are implicitly
taken into account by use of the exact scattering phase probed in the
analyticity strip.  The {\em a posteriori} success of our reasoning is
important for two reasons.  First, our construction is as simple as
the standard TBA, however avoiding the problems of dealing with density
functions for (up to) infinitely many bound states.  This may be of great
advantage in the study of more complicated systems.  Second, we have a
simple particle approach to the Heisenberg chain which will allow for a
study of transport properties like the Drude weight which has not been
possible within the path integral approach \cite{Klumper1993}.

We also demonstrate here that using our integral equations one
can improve the accuracy and extend the temperature range of
numerical calculations of $\chi(t)$ and $C(t)$ for the $S = 1/2$
Heisenberg chain {\em on the lattice} far beyond those of previous
calculations.  We find agreement of our data with the above theory
of Lukyanov \cite{Lukyanov1997} for $\chi(t)$ to high accuracy ($\lesssim
1\times 10^{-6}$) over a temperature range spanning 18 orders of
magnitude, $5\times 10^{-25}\leq t \leq 5\times 10^{-7}$; the agreement
in the lower part of this temperatures range is much better, ${\cal
O}(10^{-7})$.  For $C(t)$, the logarithmic correction in Lukyanov's
theory is insufficient to describe our numerical data accurately even
at very low $t$, so we estimate the coefficients of the next two
logarithmic correction terms in his theory from our $C(t)$ data.

\noindent
{\em Derivation of integral equations}

We start with the partially anisotropic Hamiltonian ${\cal H} = J
\sum_{i}({S}^x_i{S}^x_{i+1}+ {S}^y_i{S}^y_{i+1}+
\cos(\gamma){S}^z_i{S}^z_{i+1})-h\sum_{i}{S}^z_i$ with $0<\gamma<\pi/2$
and magnetic field $h$.  The dynamics of
the magnons, i.e.\ the elementary excitations above the ferromagnetic
state, constitute the Bethe ansatz. Momentum $p$ and energy $\epsilon$
are suitably parametrized in terms of the spectral parameter $x$
\be
p(x)=i\log\frac{\sinh(x-i\gamma/2)}{\sinh(x+i\gamma/2)}, \quad
\epsilon(x)=J\frac{\sin\gamma}{2} p'(x)-h,
\ee
where real values are obtained for Im$x=0$ and
$-\pi/2$, defining magnon bands of type ``{$+$}" and ``{$-$}".

Any two magnons with spectral parameters $x$ and $y$ scatter with phase shift
$\Theta(x-y)$ where
\be
\label{phase}
\Theta(z)=-i\log\frac{\sinh(z-i\gamma)}{\sinh(z+i\gamma)}.
\ee
Next we apply the standard TBA\cite{TakTBA}
just to the
magnons and ignore bound states! However,
the magnons on \hbox{band $-$} are considered
for spectral parameter $x$ with \hbox{Im $x$} $=-\gamma$ hence avoiding the
branch cut in the scattering phase.

The density functions for particles $\rho_j$
and holes $\rho_j^h$ for the bands $j=+$, $-$ give rise to the definition
of the ratio function $\eta_j=\rho_j^h/\rho_j$. Our analysis shows that
$\eta_+$ and $\eta_-$ are analytic continuations of each other. Quantitatively
we find $\eta_-(x+i\gamma)=\eta_+(x)=:\eta(x)$ subject to
the non-linear integral equation
\be
\log\eta(x)=\frac{\epsilon(x)}{T}+\int_C\kappa(x-y)\log(1+\eta^{-1}(y))dy
\label{NLIE}
\ee
where $\kappa(x)=\frac{1}{2\pi}\Theta'(x)$ and
$C$ is a contour consisting of the paths $C_+$ and $C_-$
with Im $y=0$ and $-\gamma$
encircled in clockwise manner.
Substituting $\log(1+\eta^{-1})=\log(1+\eta)-\log\eta$ on the contour $C_+$
and resolving for $\log\eta$ we find
\bea
\log\eta(x)=-\frac{\bar\epsilon(x)}{T}
&+&\int_{C_+}\bar\kappa(x-y)\log(1+\eta(y))dy\cr
&-&\int_{C_-}\bar\kappa(x-y)\log(1+\eta^{-1}(y))dy
\eea
with 
\bea
\bar\epsilon(x)&=&J\frac{\sin\gamma}{2}e_0(x)
+\frac{\pi}{2(\pi-\gamma)}h,\ 
e_0(x)=\frac{\frac{\pi}{\gamma}}{\cosh\frac{\pi}{\gamma}x},\\
\bar\kappa(x)&=&\frac{1}{2\pi}\int_{-\infty}^\infty\frac
{\sinh(\frac{\pi}{2}-\gamma)k}
{2\cosh\frac{\gamma}{2}k\sinh\frac{\pi-\gamma}{2}k}\e^{ikx}dk~.
\label{kernel}
\eea
Finally, ignoring $T$ and $h$ independent contributions
we obtain the free energy as
\be
f=-\frac{T}{2\pi}\int_{-\infty}^\infty  e_0(x)
\log\left[(1+\eta(x))(1+\eta^{-1}(x-i\gamma))\right] dx.
\label{EqFE}
\ee

\noindent
{\em Numerical study of low-$T$ behavior}

Lukyanov's low-$t$ asymptotic expansion of $\chi(t)$ is\cite{Lukyanov1997}
\begin{eqnarray}
\chi_{\rm lt,g}(t)J = {1\over\pi^2}\bigg\{1 &+& {g\over 2} + {3 g^3\over
32} + {\cal O}(g^4)\nonumber\\ &+& {\sqrt{3}\over \pi} t^2 [1 + {\cal
O}(g)]\bigg\}~,\label{EqsLukyanov:a}
\end{eqnarray}
where $g(t/t_0)$ obeys the transcendental equation
$
\sqrt{g}\exp({1/g}) = {t_0/ t},
$
with a unique value of $t_0$ given by
$t_0 = \sqrt{\pi/ 2}\exp(\gamma+1/4)\approx
2.866$
where $\gamma$ is \mbox{Euler's} constant.  His
expansion for the free energy per spin at $h = 0$ \cite{Lukyanov1997}
yields the specific heat per spin as
\begin{eqnarray} C_{\rm lt,g}(t) = {2 t\over 3 }\Big[1 &+& {3\over 8}\,g^3
+ {\cal O}(g^4)\Big]\nonumber\\
\nonumber\\ &+& {(2)3^{5/2}t^3\over 5\pi}[1 + {\cal O}(g)]~,
\label{EqCLusnikov}
\end{eqnarray}
where the exact prefactor $2t/3$ was found by Affleck in
1986\cite{Affleck1986}, and the prefactor 3/8 in the logarithmic
correction term agrees with \cite{Klumper1998,Affleck1989,Karbach95}.

\begin{figure}
\epsfxsize=3in
\centerline{\epsfbox{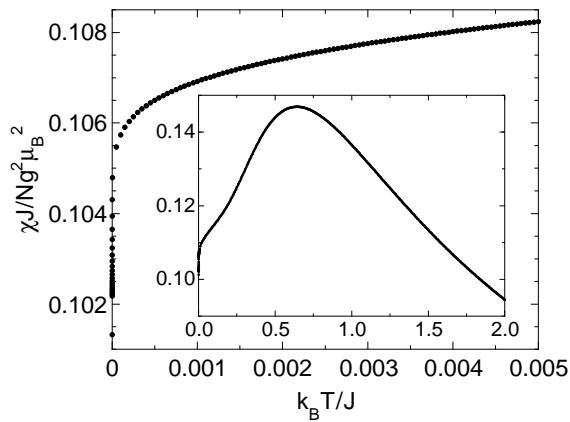}}
\vglue 0.1in 
\caption{Magnetic susceptibility $\chi$ at low temperature $T$ for the
  spin $S = 1/2$ antiferromagnetic uniform Heisenberg chain. In the
  inset $\chi(T)$ is shown on a larger temperature scale.}
\label{Fig1}
\end{figure}

Numerical data for $\chi(t)$ and $C(t)$ were obtained using
our free energy expression~(\ref{EqFE}).  
These data are considerably more accurate than those presented previously
in \cite{Klumper1998}.
Our $\chi(t)J$ data, and the
exact value $1/\pi^2$ at $t = 0$ \cite{Griffiths1964}, are plotted in
Fig.~\ref{Fig1}.  The calculations have an absolute accuracy of $\approx
1\times 10^{-9}$.  The data show a maximum at a temperature $t^{\rm max} =
0.6\,408\,510(4)$ with a value $\chi^{\rm max}J = 0.146\,926\,279(1)$,
yielding the $J$-independent product $\chi^{\rm max}T^{\rm max} =
0.0\,941\,579(1)$.  These values are consistent within the errors with
those found by Eggert, Affleck and Takahashi\cite{Eggert1994}, but are
much more accurate.

\begin{figure}
\epsfxsize=2.8in
\centerline{\epsfbox{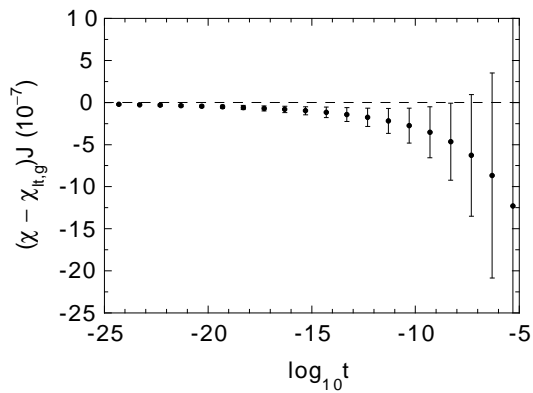}}
\vglue 0.1in
\caption{Semilog plot vs temperature $t$ at low $t$ of the difference
between our Bethe ansatz magnetic susceptibility $\chi J$ data and the
prediction $\chi_{\rm lt,g}J$ of Lukyanov's
theory\protect\cite{Lukyanov1997}.  The error bars are the estimated
uncertainties in $\chi_{\rm lt,g}(t)J$.}
\label{Fig2}
\end{figure}

The differences between our low-$t$ Bethe ansatz $\chi(t)J$ calculations
and Lukyanov's theoretical $\chi_{\rm lt,g}(t)J$ prediction in
Eq.~(\ref{EqsLukyanov:a}) are shown in Fig.~\ref{Fig2}.  The
error bar on each data point is the estimated uncertainty in $\chi_{\rm
lt,g}J$ arising from the presence of the unknown ${\cal O}(g^4)$ and
higher-order terms in Eq.~(\ref{EqsLukyanov:a}), which was arbitrarily set
to $g^4(t)/\pi^2$; the uncertainty in the $t^2$ contribution,
$\sim\sqrt{3}t^2g(t)/\pi^3$, is negligible at low $t$ compared to this.
At the lower temperatures, the data agree extremely well with the
prediction of Lukyanov's theory.  At the highest temperatures,
higher order $t^n$ terms also become important.  Irrespective of these
uncertainties in the theoretical prediction at high temperatures, we can
safely conclude directly from Fig.~\ref{Fig2} that our numerical
$\chi(t)$ data are in agreement with the theory of
Lukyanov\cite{Lukyanov1997} to within an absolute accuracy of $1\times
10^{-6}$ (relative accuracy $\approx 10$ ppm) from $t = 5\times 10^{-25}$
to $t = 5\times 10^{-7}$.  The agreement at the lower temperatures,
${\cal O}(10^{-7})$, is much better than this.

\begin{figure}
\epsfxsize=3in
\centerline{\epsfbox{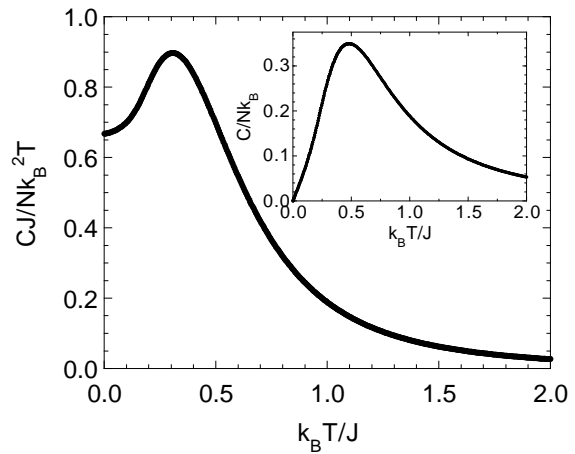}}
\vglue 0.1in
\caption{Electronic specific heat coefficient $C/T$
  versus temperature $T$ for the $S = 1/2$ AF uniform Heisenberg chain.
  In the inset the specific heat $C$ versus $T$ is shown.}
\label{Fig3}
\end{figure}

\begin{figure}
\epsfxsize=3in
\centerline{\epsfbox{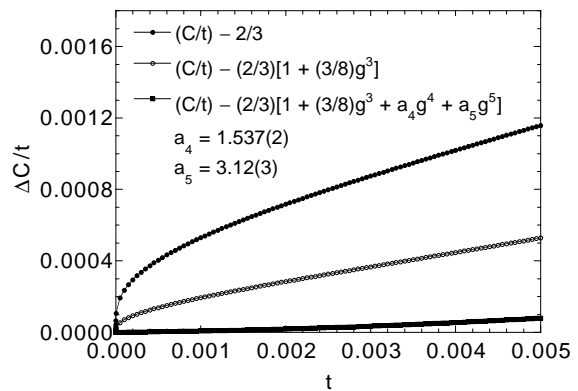}}
\vglue 0.1in
\caption{Difference $\Delta C/t$ between our Bethe ansatz electronic
specific heat coefficient data and the exact coefficient 2/3 at $t =
0$ (top data set), versus temperature $t$.  Successive data sets show the
influence of subtracting cumulative logarithmic correction terms.}
\label{Fig4}
\end{figure}

Our $C(t)$ data for $t\leq 2$ are shown in the inset of
Fig.~\ref{Fig3} and have an
estimated accuracy of $3\times 10^{-10}C(t)$. The data show a
maximum with a value $C^{\rm max} = 0.3\,497\,121\,235(2)$ at a
temperature $t_C^{\rm max}= 0.48\,028\,487(1)$.  The electronic specific
heat coefficient $C(t)/t$ is plotted in Fig.~\ref{Fig3}.  These data
exhibit a maximum with a value $(C/t)^{\rm max}= 0.8\,973\,651\,576(5)$ at
$t_{\rm C/t}^{\rm max}= 0.30\,716\,996(2)$.  The existence of low-$t$ log
corrections to $C(t)$ is revealed in the top plot of $\Delta C(t)/t$ in
Fig.~\ref{Fig4}, where $\Delta C(t) = C(t) - 2 t/3$ and $2t/3$ is the
low-$t$ limit of $C(t)$.  The influence of the $g^3$ log correction term
in Eq.~(\ref{EqCLusnikov}) is evaluated by subtracting it in the plot of
$\Delta C(t)/t$ as shown by the middle curve in Fig.~\ref{Fig4}.  The $t
= 0$ singularity is still present but with reduced amplitude; this
demonstrates that additional logarithmic correction terms are important
within the accuracy of the data.

\begin{figure}
\epsfxsize=2.8in
\centerline{\epsfbox{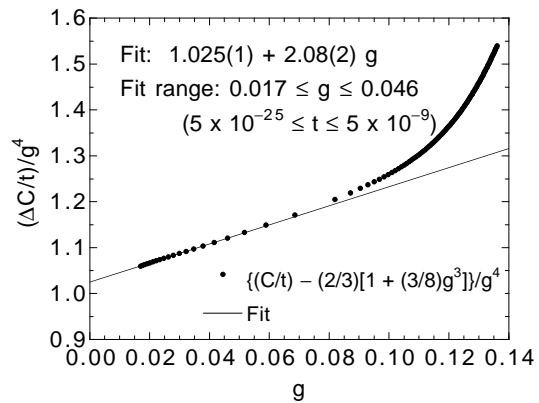}}
\vglue 0.1in
\caption{Plot showing the estimation of the coefficients of the $g^4$ and
$g^5$ logarithmic correction terms in Eq.~(\protect\ref{EqCLusnikov}).
The error bars on the data points are smaller than the symbols.}
\label{Fig5}
\end{figure}

We estimate the unknown coefficients of the next two logarithmic
correction ($g^4,\ g^5$) terms in Eq.~(\ref{EqCLusnikov}) from our $C(t)$
data as follows.  From Eq.~(\ref{EqCLusnikov}), if we plot the data as
$[C(t)/t - (2/3)(1+3g^3/8)]/g^4$ vs $g$ and fit the lowest-$t$ data by a
straight line, the $y$-intercept gives the coefficient of the
$g^4$ term and the slope gives the coefficient of the $g^5$ term.  We
fitted a straight line to the data in such a plot for $5\times10^{-25}\leq
t\leq 5\times10^{-9}$ as shown by the weighted linear fit in
Fig.~\ref{Fig5} where the parameters of the fit are given in the figure.
By subtracting the influences of these two logarithmic correction terms
from the middle data set as shown in the bottom data set in
Fig.~\ref{Fig4}, the singular behavior as $t\to 0$ is largely removed,
leaving a behavior which is close to a $t^2$ dependence as predicted by
the last term in Eq.~(\ref{EqCLusnikov}).  Further discussion of the
predictions of \cite{Lukyanov1997}, and high-accuracy fits
($0\leq t\leq 5$) to our $C(t)$ and $\chi(t)$ data and the respective
exact $t = 0$ values, will be presented elsewhere\cite{Johnston1999}.

In conclusion, we have presented an analytic approach to
the thermodynamics of the $S = 1/2$ AF Heisenberg chain on the basis of a
finite number of elementary excitations.  We envisage that this approach
can be generalized to study a variety of other systems such as Hubbard
and $t$-$J$ models, quantum spin chains with higher symmetries and
systems with orbital degrees of freedom.  Our free energy expression has
allowed numerical calculations of $\chi(t)$ and $C(t)$ for the Heisenberg
chain to be carried out to much higher accuracy and to much lower
temperatures than heretofore attained.  Our $\chi(t)$ data are in
excellent agreement with the theory of Lukyanov\cite{Lukyanov1997} at low
$t$.  The logarithmic correction in Lukyanov's theory for $C(t)$ is found
insufficient to describe our $C(t)$ data accurately even at very low $t$.
However, the $t$ dependence of the deviation agrees with the form of his
theory, which enabled us to estimate the unknown coefficients of the next
two logarithmic correction terms in his theory for $C(t)$ from our $C(t)$
data.  Thus we have verified Lukyanov's theory\cite{Lukyanov1997} of a
critical system perturbed by marginal operators and have given evidence
that his asymptotic expansion can be systematically extended to higher
order.

The authors acknowledge valuable discussions with U.~L\"ow and K.
Fabricius.  Comparison of our results with their numerical data for
the thermodynamics of finite systems proved essential to achieve high
accuracy in the treatment of the nonlinear integral equations.  D.C.J.
thanks the University of Cologne and the Stuttgart Max-Planck-Institut
f\"ur Festk\"orperforschung for their hospitality.  A.K.
acknowledges financial support by the {\it Deutsche
Forschungsgemeinschaft} under grant No.~Kl~645/3 and by the research
program of the Sonderforschungsbereich 341, K\"oln-Aachen-J\"ulich.  Ames
Laboratory is operated for the U.S. Department of Energy by Iowa State
University under Contract No.\ W-7405-Eng-82.  The work at Ames
was supported by the Director for Energy Research, Office of
Basic Energy Sciences.

\begin{references}

\bibitem{Bethe31} H. A. Bethe, Z. Phys.\ {\bf 71}, 205 (1931).

\bibitem{TakTBA} M. Takahashi, Prog.\ Theor.\ Phys.\ {\bf 46}, 401
(1971); {\it ibid.}~{\bf 50}, 1519 (1973).

\bibitem{SuzukiI87} M. Suzuki and M. Inoue, Prog.\ Theor.\ Phys.\ {\bf
78}, 787 (1987).

\bibitem{Tak91} M. Takahashi, Phys.\ Rev.\ B {\bf 43}, 5788 (1991); {\it
ibid.}\ {\bf 44}, 12\,382 (1991).

\bibitem{Eggert1994}S. Eggert, I. Affleck, and M. Takahashi, Phys.\ Rev.\
Lett.\ {\bf 73}, 332 (1994).

\bibitem{Bonner1964}J. C. Bonner and M. E. Fisher, Phys.\ Rev.\
{\bf 135}, A640 (1964).

\bibitem{Lukyanov1997}S. Lukyanov, Nucl.\ Phys.\ B {\bf 522}, 533 (1998).

\bibitem{Klumper1993}A. Kl\"umper, Z. Phys. B {\bf 91}, 507 (1993).

\bibitem{Klumper1998}A. Kl\"umper, Eur.\ Phys.\ J. B {\bf 5}, 677 (1998).

\bibitem{Affleck1986}I. Affleck, Phys.\ Rev.\ Lett.\ {\bf 56}, 746 (1986).

\bibitem{Affleck1989}I. Affleck, D. Gepner, H. J. Schulz, and T. Ziman,
J.~Phys.\ A {\bf 22}, 511 (1989).

\bibitem{Karbach95}M. Karbach and K.-H. M\"utter, J. Phys.\ A {\bf 28},
4469 (1995).

\bibitem{Griffiths1964}R. B. Griffiths, Phys.\ Rev.\ {\bf 133}, A768
(1964); C. N. Yang and C. P. Yang, Phys.\ Rev.\ {\bf 150}, 327 (1966).

\bibitem{Johnston1999}D. C. Johnston {\it et al.}, unpublished.

\end{references}
\end{document}